# Flocking of V-shaped and Echelon Northern Bald Ibises with Different Wingspans: Repositioning and Energy Saving

Amir Mirzaeinia[1], Mehdi Mirzaeinia[2], Mostafa Hassanalian[3]*

**V-shaped and echelon formations help migratory birds to consume less energy for migration. As the case study, the formation flight of the Northern Bald Ibises is considered to investigate different effects on their flight efficiency. The effects of the wingtip spacing and wingspan are examined on the individual drag of each Ibis in the flock. Two scenarios are considered in this study, (1) increasing and (2) decreasing wingspans toward the tail. An algorithm is applied for replacement mechanism and load balancing of the Ibises during their flight. In this replacing mechanism, the Ibises with the highest value of remained energy are replaced with the Ibises with the lowest energy, iteratively. The results indicate that depending on the positions of the birds with various sizes in the flock, they consume a different level of energy. Moreover, it is found that also small birds have the chance to take the lead during the flock.**

## I. Introduction

There are 1800 species of birds that migrate through seasonal long-distance routes from hundreds up to several thousand kilometers[1]. For example, Bar-tailed godwits as one of the non-stop migratory birds travel for 11600 km[1,2]. Canada Geese with an average speed of about 18 m/s migrate for 3,200 to 4,800 Km in a single day if the weather is good[44]. The Arctic terns migrate from the Arctic to the Antarctic[1]. Besides the mentioned migratory birds, other bird species including pelicans, swans, cormorants, cranes, and Ibis migrate in an echelon or V-shaped formations[1]. The formation flight of migrating birds has been studied for more than fifty years. This behavior of the migrating birds has increased the interest of drones' designers to come with some new solutions for enhancing the drones' endurance and efficiency[3-12].

One of the interesting birds that apply echelon flight formation during their migration is Northern bald Ibises. In 2014, Portugal et al. showed that Northern bald Ibises which performing an echelon or V-shaped flocking, locate themselves in aerodynamically optimum positions[13]. In this study, they suggested that Northern bald Ibises in V-shaped formation have phasing strategies to cope with the dynamic wakes produced by flapping wings[13]. Voelkl et al. in 2015 studied the flight behavior of a flock of juvenile Northern bald Ibis during autumn migration. They indicated that the leading time of a bird in a formation flight is strongly correlated with the time that it can itself benefit from flying in the wake of another bird. They showed that on the dyadic level, birds generally match the time they spend in the wake of each other by frequent switching of the leading position. Moreover, they found evidence that the animals' tendency to reciprocate in leading has a substantial influence on the size and cohesion of the flight formations[14]. In 2015, Barlein et al.[15] examined experimentally the physiological changes and energy consumption in Northern bald Ibis during their flights[15]. The results of this study indicated that balancing aerobic and anaerobic metabolism generates high amounts of energy[15]. In 2017, Voelkl and Fritz analyzed the formation flight of Northern bald Ibis on their first southbound migration. In this work, they found that behavioral adaptations for dealing with physiological constraints on long-distance migrations require the formation of social groups with different characteristics[1].

---

[1] PhD. student, Department of Computer Science and Engineering, New Mexico Tech, Socorro, NM 87801, USA.
[2] M.Sc. student, Department of Electrical Engineering, Amir Kabir University of Technology, Tehran, Iran.
[3] Assistant Professor, Department of Mechanical Engineering, New Mexico Tech, Socorro, NM 87801, USA.

These researches have indicated that juvenile Northern bald Ibis can play a positive role during the flocking by precisely matching times they spend in the advantageous and disadvantageous positions and taking turns[8,15]. Moreover, it has been shown that the tendency of these migrating birds to lead the flock has a significant effect on the consistency and size of the formations. As shown before, the birds in an echelon or V-shaped formation undergo to a different value of drag forces. It has been demonstrated that birds in the leading and tail positions are consuming more energy than the birds in the middle.

Mirzaeinia et al.[8] in 2019 studied the energy saving of echelon flocking Northern bald Ibises with variable wingtips spacing[8]. They showed that the wingtip spacing can play an important role in the energy saving of these migratory birds. Mirzaeinia and Hassanalian[10] in 2019 also presented an aerodynamics analysis of the V-shaped flocking Canada geese and their energy-saving consequences because of the repositioning. This analysis showed that how the Canada geese can fly very far in a single day through repositioning. Extensive analysis indicated that leader and tail positions switching of fourteen Canada geese can improve the flight range and endurance of these migrating birds more than 44.5%[10].

In both studies conducted by Mirzaeinia and Hassanalian[8,10], it was assumed that all the birds have a similar wingspan, weight, and speed. They investigated the effects of replacing the lead and tail birds with middle ones. Their studies uncovered that the replacement scenario help to balance the drag load during the flight and increase the flight time and distance consequently. However, they analyzed the V-formation flight of the same size which is not a realistic configuration since birds have a range of dimensions. Birds with variable sizes in a V-shaped formation flight may have different performance. Moreover, the position of birds with different sizes may have different drag loads. For example, if the larger birds stay at the front and the younger bird follow them at the backside or the other way around. In this paper, the formation flight of the Northern bald Ibises is studied for birds with different wingspans and weights. Two different case scenarios are presented in this paper. In the first scenario, it is assumed that the wingspans of the birds decreasing toward the tail and in the second scenario, the wingspans increase toward the tail position.

## II. Flight Characteristics and Geometric of the Northern Bald Ibises

The Northern bald Ibises have a migratory pattern that at its' largest route is from Syria to central Ethiopia; a distance of 2,816 Km[16]. These birds have a length of 70-80, a wingspan of 125-140 cm, a weight of 1.0 kg to 1.3 kg, and an average flight speed of 14 m/s[17,18]. The Northern Ibises generally migrate in an echelon or V-shaped formations [8]. These birds can save more energy during the migration by applying these formations. The flight altitude for the Northern Ibis changes from 60 m to 224 m above ground when migrating. The Northern Ibis can migrate up to 500 Km per day when conditions are favorable[18]. In Figures 1(a) and 1(b), views of these migratory birds and their echelon formation are shown.

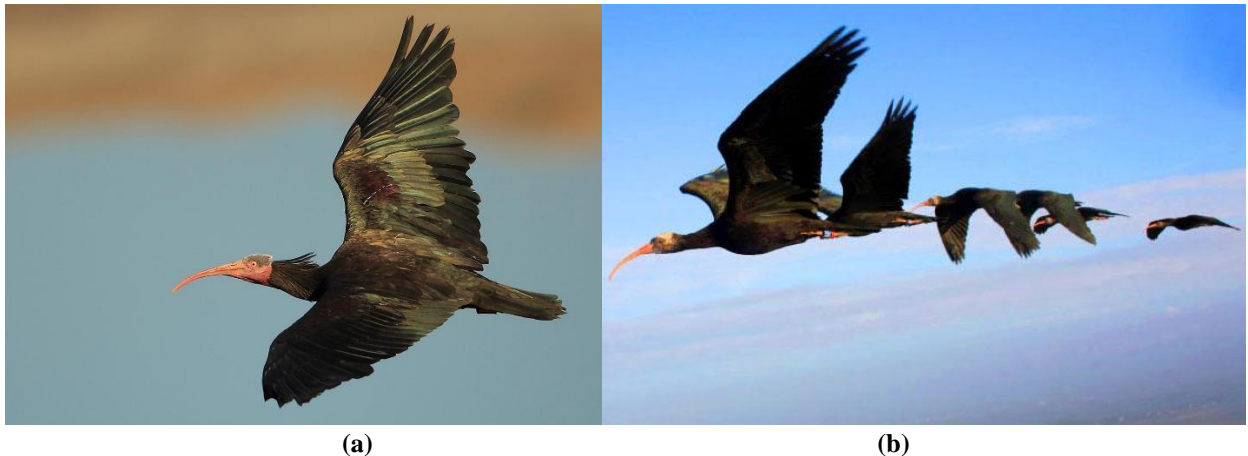

(a)      (b)

**Figure 1. Views of (a) Northern bald Ibis and (b) echelon formation.**

## III. Drag Modeling of Echelon Flocking Ibises with Different Sizes

During the echelon or V-shaped formation of Northern bald Ibises, the high-pressure air under the wings flows around the tips to a region of low air pressure above the wings. This flow of air creates two vortices that form a central region of downwash immediately behind the Ibises and upwash regions outboard of their wings. These generated upwashes provide an extra lift for the following Ibises and consequently reduce their induced drag. In this study, it is assumed that wings have a fixed structure. Applying the modeling developed and modified by Mirzaeinia and Hassanalian, the total drag of flocking Northern bald Ibises is calculated as bellows[8]:

$$D_I(n) = \sum_{i=1}^{n}\sum_{j=1}^{n} \frac{(D_{Iii}+D_{Ijj})}{\pi^2} \log\left[1-\left(\frac{a_i+a_j}{\frac{b_i+b_j}{2}+\sum_{k=i+1}^{j-1} b_k + \sum_{k=i}^{j-1} s_k}\right)^2\right] \quad (1)$$

where $D_I$ is the induced drag for the wings of the Ibises calculated by the following equation, $b$ is the wingspan, $s$ is the wingtip spacing, and $a$ is the half of the distance between the tip vortices[8].

$$D_I = 2a\rho\Gamma w_d = \frac{2(mg)^2}{\pi\rho b^2 V^2} \quad (2)$$

where $\Gamma$ is the strength of the vortex, $w_d$ is the downward velocity, $V$ is the flight speed, $\rho$ is the air density, and $m$ and $g$ are mass and gravity acceleration of the Ibises, respectively. In this study, it is assumed that the ratio of $a$ to $b$ is equal to $\pi/8$. The total drag of each individual Ibis (Ibis-$k$) considering echelon formation is given by:

$$D_{Ik} = D_{Ikk} + \frac{1}{\pi^2}\sum_{j=1}^{n}(D_{Ikk}+D_{Ijj}) \log\left[1-\left(\frac{a_k+a_j}{\delta y_{kj}}\right)^2\right] \quad (3)$$

where $D_{Iii}$ and $D_{Ijj}$ are the self-induced drags of Ibis-$i$ and Ibis-$j$, respectively, and $a_i$ and $a_j$ are the distance between the tip vortices and root chord of each Ibis. $\delta y_{ij}$ is expressed as:

$$\delta y_{ij} = \frac{b_i+b_j}{2} + \sum_{k=i+1}^{j-1} b_k + \sum_{k=i}^{j} s_k \quad (4)$$

Badgerow and Hainsworth in 1981 showed that a bird gains the greatest benefits if its wingtip overlaps laterally with that of the bird in front. They proposed that the optimal wingtip spacing can be calculated as below[19]:

$$s_{opt} = 0.5b(0.78-1) = -0.11b \quad (5)$$

In this study, the corresponding $s$ for each bird with a different wingspan is considered. In the previous study conducted by Mirzaeinia and Hassanalian[8], it was shown that the wingtip spacing has a considerable effect on the drag of individual Ibises with the same wingspan in a flock.

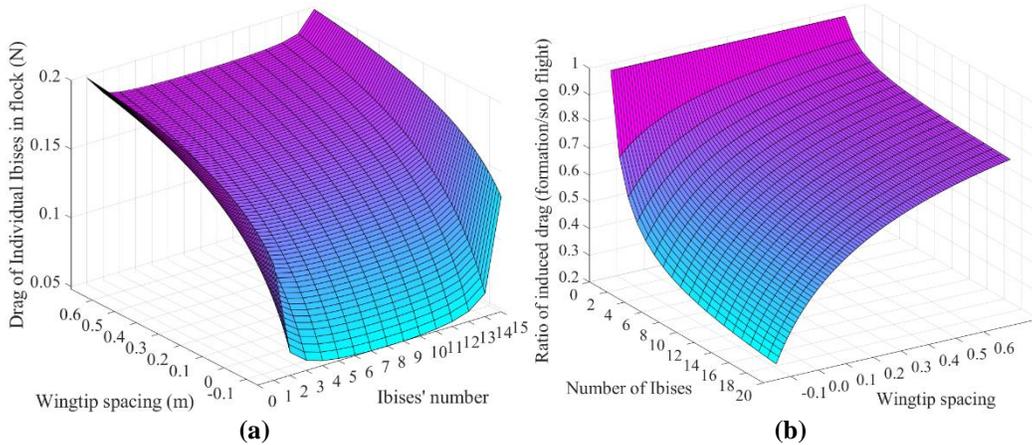

**Figure 2. (a) Drag of individual Ibises and (b) ratio of induced drag for 15 Ibises in different wingtip spacing.**

## IV. Individual Drag of Northern Bald Ibises with Variable Sizes in Echelon Formation

Different scenarios are investigated in this study. It is assumed that the Northern Bald Ibises have a wingspan from 1.25 m to 1.40 m with an increment of 0.01 m and they have a corresponding weight from 1 kg to 1.3 kg. The speed of all the Ibises is considered constant equal to 14 m/s. In Figures 3(a) and 3(b), different proposed scenarios are shown.

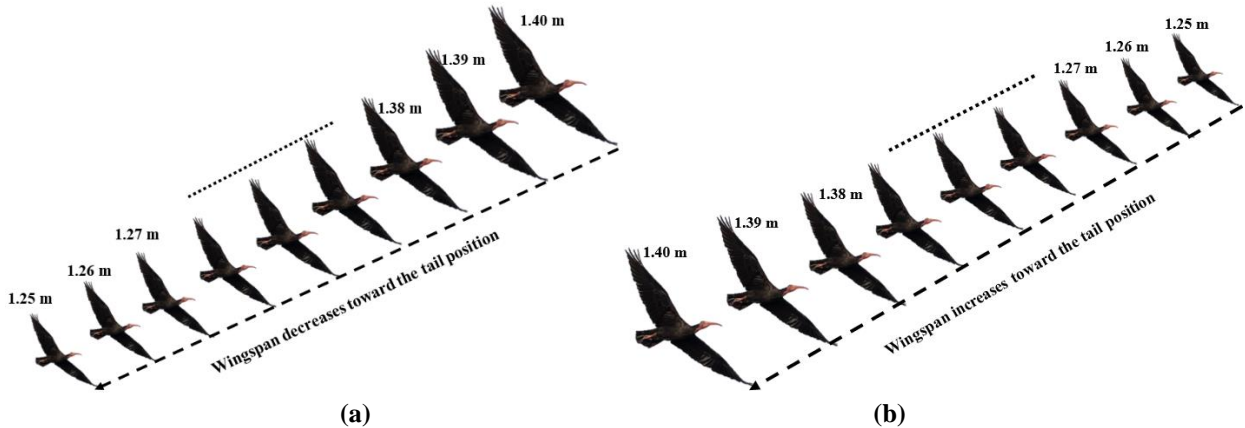

(a)                (b)

**Figure 3. Views of flocks of Ibises with variable wingspans (a) wingspan decreases towards the tail position (b) wingspan increases towards the tail position.**

In Figure 4, the drag of each Ibis in echelon flight is shown. As can be seen, the birds that have higher values of wingspans in each scenario generate more drag. It can be concluded from Figure 4 that the leader and the tail Ibises are consuming more energy than the rest of the Ibises during the flocking flight.

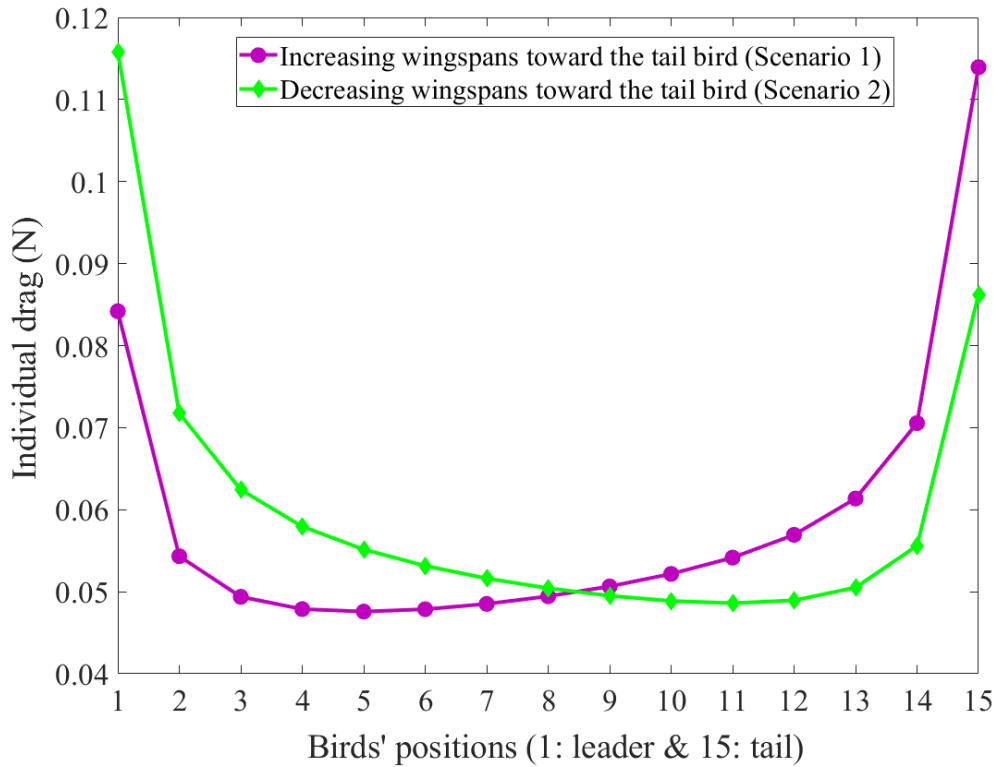

**Figure 4. Drag for individual Ibis in flocking flight for two different scenarios.**

In Table 1, the characteristics of the considered Ibises in two discussed scenarios (wingspan decreases towards the tail position, and wingspan increases towards the tail position) are shown.

Table 1. Considered geometric characteristics for Ibises in the flock.

| Position Number | Scenario 1 | | | Scenario 2 | | |
|---|---|---|---|---|---|---|
| | Wingspan (m) | Mass (kg) | Drag (N) | Wingspan (m) | Mass (kg) | Drag (N) |
| 1 | 1.25 | 1.00 | 0.0842 | 1.39 | 1.28 | 0.1158 |
| 2 | 1.26 | 1.02 | 0.0543 | 1.38 | 1.26 | 0.0718 |
| 3 | 1.27 | 1.04 | 0.0494 | 1.37 | 1.24 | 0.0624 |
| 4 | 1.28 | 1.06 | 0.0479 | 1.36 | 1.22 | 0.0580 |
| 5 | 1.29 | 1.08 | 0.0476 | 1.35 | 1.20 | 0.0552 |
| 6 | 1.30 | 1.10 | 0.0479 | 1.34 | 1.18 | 0.0532 |
| 7 | 1.31 | 1.12 | 0.0485 | 1.33 | 1.16 | 0.0516 |
| 8 | 1.32 | 1.14 | 0.0495 | 1.32 | 1.14 | 0.0504 |
| 9 | 1.33 | 1.16 | 0.0507 | 1.31 | 1.12 | 0.0495 |
| 10 | 1.34 | 1.18 | 0.0522 | 1.30 | 1.10 | 0.0489 |
| 11 | 1.35 | 1.20 | 0.0542 | 1.29 | 1.08 | 0.0486 |
| 12 | 1.36 | 1.22 | 0.0569 | 1.28 | 1.06 | 0.0490 |
| 13 | 1.37 | 1.24 | 0.0614 | 1.27 | 1.04 | 0.0505 |
| 14 | 1.38 | 1.26 | 0.0706 | 1.26 | 1.02 | 0.0556 |
| 15 | 1.39 | 1.28 | 0.1139 | 1.25 | 1.00 | 0.0862 |

In Figure 5, the bar graph for the individual drag of each Ibis with a specified wingspan in considered scenarios is shown. As can be seen in Figure 5, besides the wingspan, the position of the birds in the flock plays an important role in the value of the drag. The Ibises with the maximum and minimum wingspans that are located in the tail and lead positions have the highest drag compared to other birds.

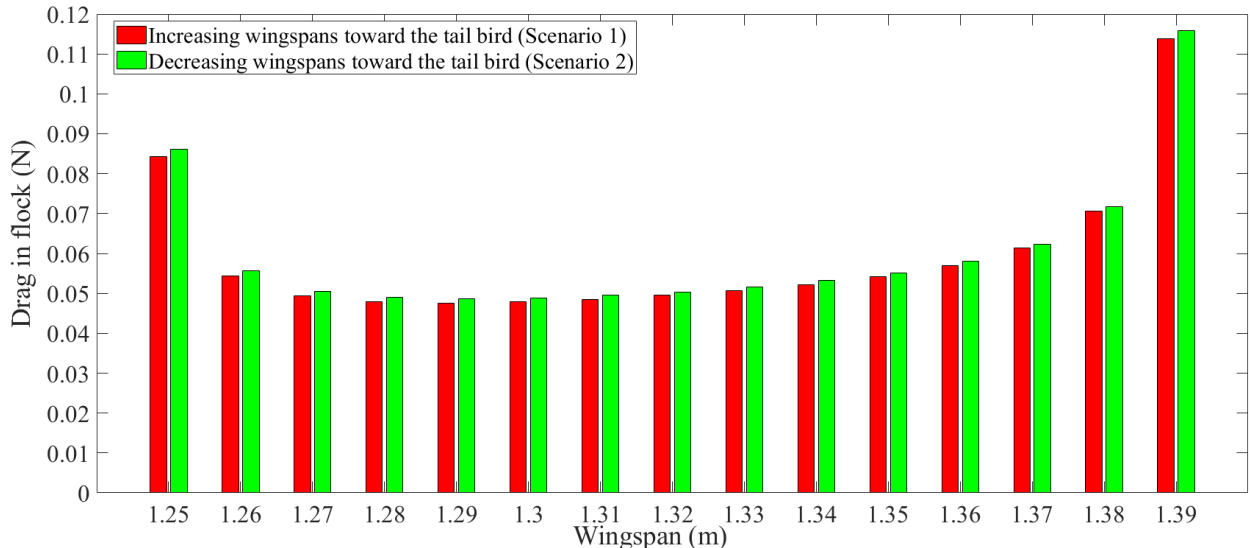

**Figure 5. Drag of individual Ibises with a specified wingspan for two different scenarios.**

### V. Northern Bald Ibises and Reformation

In this paper, we shuffle the birds in each replacement step. However, different types of shuffling are possible to implement. Besides, finding the optimum shuffling method is not easy to find. Therefore, the shuffle is proposed in this paper is to replace the highest-remained energy bird with the lowest-remained energy, iteratively. Algorithm 1 shows this process in more detail.

| Algorithm 1: Replacement mechanism |
|---|
| 1:     **Function** Shuffle (a, s, b, N, v, init_energy) |
|         d_ii = (2 * m^2 * 9.8 ^ 2) / (pi * 1.225 * b ^ 2 * v ^ 2); % induced drag of Ibises |
| 2:     D_total = Drag_Calculation_function (a,s,b,N,d_11); % Vector contains drag of all Ibises |
| 3:     **While** (min ( Remind energy) > 0.01 * initial energy) |
| 4:         Distance = 0.5 * Remind Energy of the leader/ D_total (1) |
| 5:         Remind Energy= Remind Energy - Distance * D_total    % Vector contains remind energy of all Ibises |
| 6:         for i=1:floor(length(b)/2) |
| 7:             max_ind = find(maximum remained energy); |
| 8:             min_ind = find(minimum remained energy); |
| 9:             Swap (max_ind, min_ind) |
| 10:        End |
| 11:     **End Function** |

Based on the above algorithm, the repositionings are carried out for two considered scenarios. In Figures 6(a) to 6(d), different steps of the replacements are shown for the initial flock with increasing wingspan toward the tail.

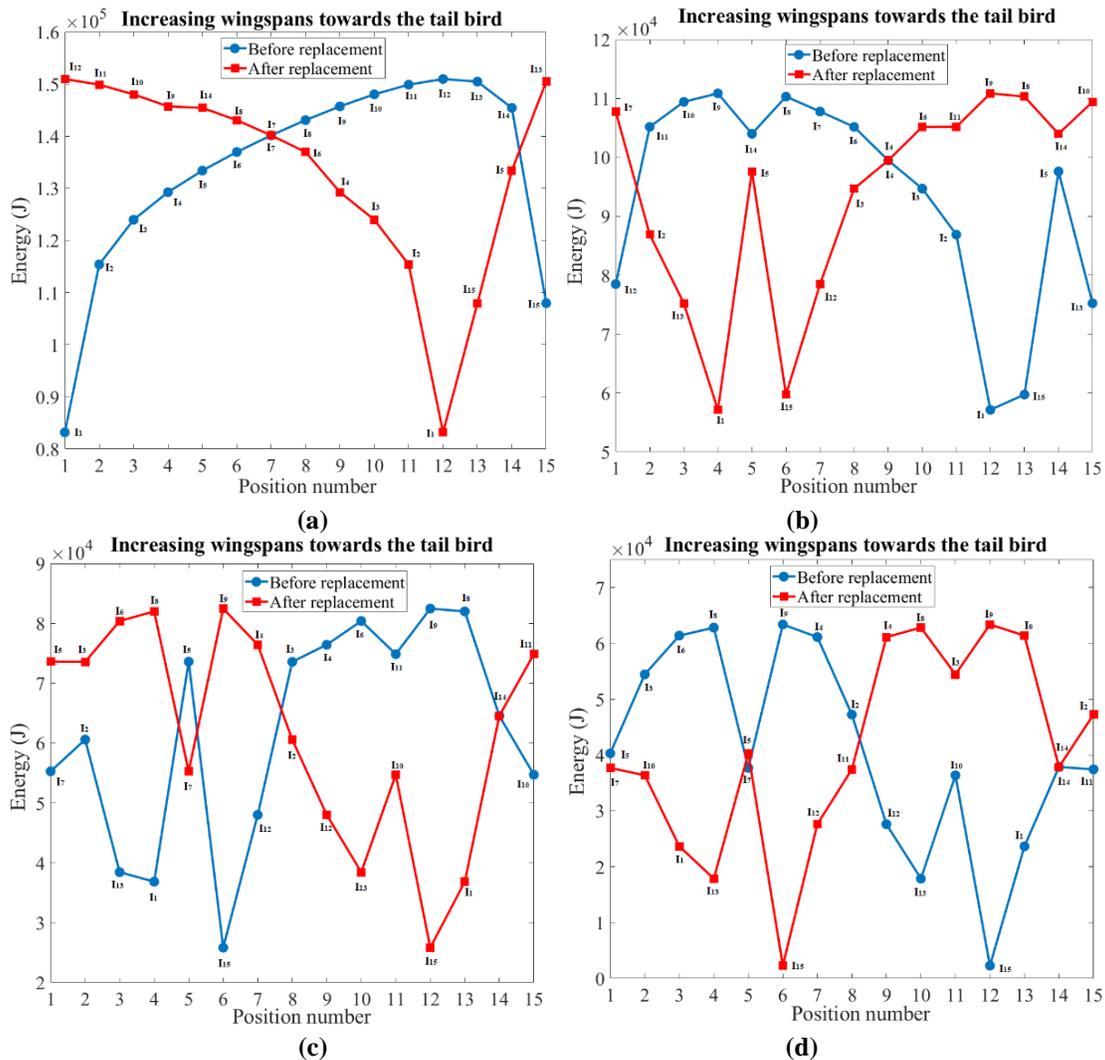

**Figure 6. Remained energy of individual Ibis before and after replacements (scenario 2).**

In this scenario, the percentages of remained energy for each Ibis are indicated in Figure 7 and Table 2 for different steps of replacement.

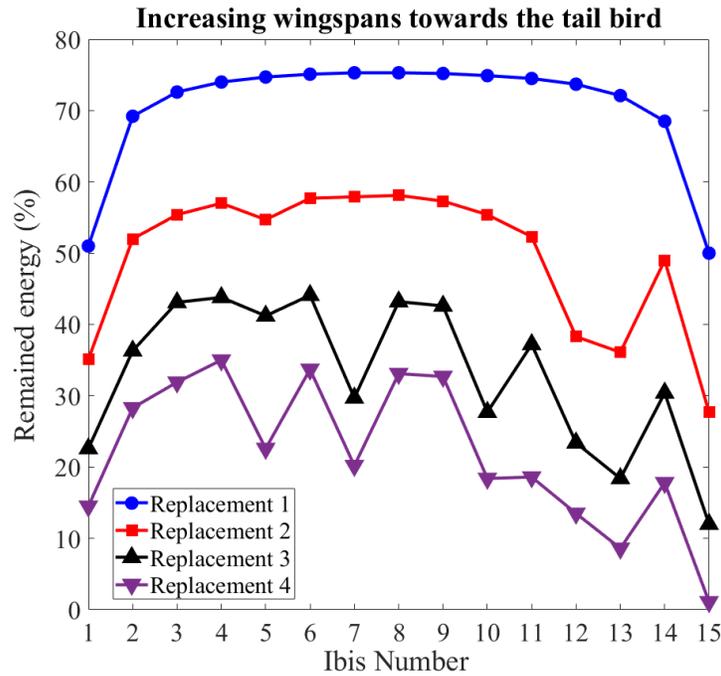

**Figure 7. Percentages of remained energy of individual Ibis after four replacements (scenario 1).**

**Table 2. Percentage of remained energy for Ibises in the flock (scenario 1).**

| Position | | Replacement 1 | | Replacement 2 | | Replacement 3 | | Replacement 4 | |
|---|---|---|---|---|---|---|---|---|---|
| Initial | Energy× $10^5$ | Ibis number | Energy (%) | Ibis number | Energy (%) | Ibis number | Energy (%) | Ibis number | Energy (%) |
| 1 | 1.6297 | 12 | 73.7 | 7 | 57.9 | 5 | 41.2 | 7 | 20.2 |
| 2 | 1.6688 | 11 | 74.5 | 2 | 52.0 | 3 | 43.1 | 10 | 18.4 |
| 3 | 1.7077 | 10 | 74.9 | 13 | 36.1 | 6 | 44.1 | 1 | 14.5 |
| 4 | 1.7464 | 9 | 75.2 | 1 | 35.1 | 8 | 43.2 | 13 | 08.6 |
| 5 | 1.7849 | 14 | 68.5 | 5 | 54.7 | 7 | 29.7 | 5 | 22.6 |
| 6 | 1.8232 | 8 | 75.3 | 15 | 27.7 | 9 | 42.6 | 15 | 01.1 |
| 7 | 1.8614 | 7 | 75.3 | 12 | 38.3 | 4 | 43.8 | 12 | 13.5 |
| 8 | 1.8993 | 6 | 75.1 | 3 | 55.4 | 2 | 36.3 | 11 | 18.6 |
| 9 | 1.9371 | 4 | 74.0 | 4 | 57.0 | 12 | 23.4 | 4 | 35.0 |
| 10 | 1.9747 | 3 | 72.6 | 6 | 57.7 | 13 | 18.4 | 8 | 33.1 |
| 11 | 2.0120 | 2 | 69.2 | 11 | 52.3 | 10 | 27.7 | 3 | 31.9 |
| 12 | 2.0492 | 1 | 51.0 | 9 | 57.3 | 15 | 12.0 | 9 | 32.7 |
| 13 | 2.0861 | 15 | 50.0 | 8 | 58.1 | 1 | 22.6 | 6 | 33.7 |
| 14 | 2.1229 | 5 | 74.7 | 14 | 49.0 | 14 | 30.4 | 14 | 17.8 |
| 15 | 2.1594 | 13 | 72.1 | 10 | 55.4 | 11 | 37.2 | 2 | 28.3 |

A similar algorithm is employed for the repositionings of Ibises that have been initially located in the flock based on decreasing the wingspan toward the tail. In Figures 8(a) to 8(d), different steps of the replacements for this scenario are indicated.

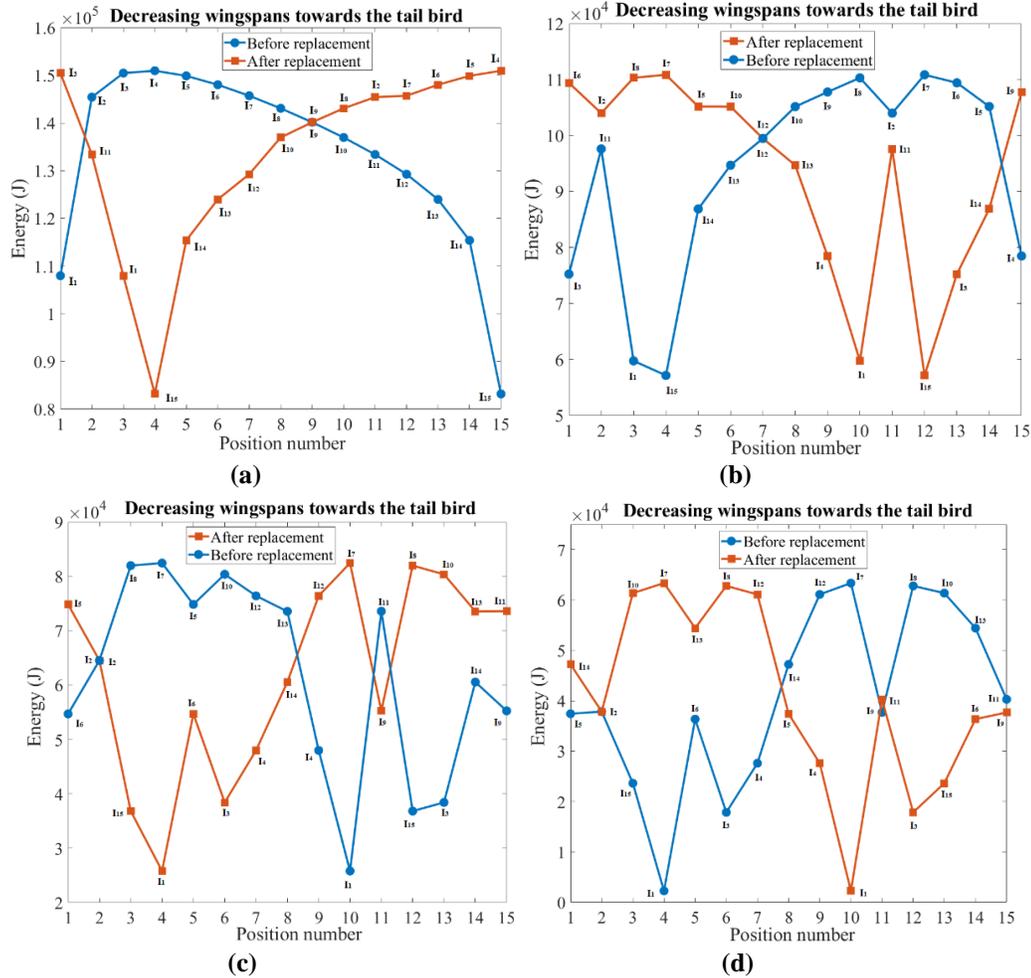

**Figure 8. Remained energy of individual Ibis before and after replacements (scenario 2).**

In Figure 9 and Table 3, the percentages of remained energy for each Ibis in echelon formation are shown for different steps of replacement. It can be seen that in this scenario, birds with different wingspans will have different levels of remained energy.

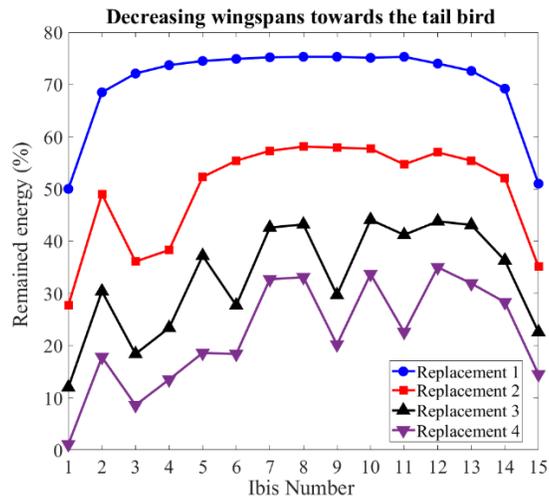

**Figure 9. Percentages of remained energy of individual Ibis after four replacements (scenario 2).**

Table 3. Percentage of remained energy for Ibises in the flock (scenario 2).

| Position | Replacement 1 | | Replacement 2 | | Replacement 3 | | Replacement 4 | |
|---|---|---|---|---|---|---|---|---|
| Initial | Energy× $10^5$ | Ibis number | Energy (%) | Ibis number | Energy (%) | Ibis number | Energy (%) | Ibis number | Energy (%) |
| 1 | 2.1594 | 3 | 72.1 | 6 | 55.4 | 5 | 37.2 | 14 | 28.3 |
| 2 | 2.1229 | 11 | 75.3 | 2 | 49.0 | 2 | 30.4 | 2 | 17.8 |
| 3 | 2.0861 | 1 | 50.0 | 8 | 58.1 | 15 | 22.6 | 10 | 33.7 |
| 4 | 2.0492 | 15 | 51.0 | 7 | 57.3 | 1 | 12.0 | 7 | 32.7 |
| 5 | 2.0120 | 14 | 69.2 | 5 | 52.3 | 6 | 27.7 | 13 | 31.9 |
| 6 | 1.9747 | 13 | 72.6 | 10 | 57.7 | 3 | 18.4 | 8 | 33.1 |
| 7 | 1.9371 | 12 | 74.0 | 12 | 57.0 | 4 | 23.4 | 12 | 35.0 |
| 8 | 1.8993 | 10 | 75.1 | 13 | 55.4 | 14 | 36.3 | 5 | 18.6 |
| 9 | 1.8614 | 9 | 75.3 | 4 | 38.3 | 12 | 43.8 | 4 | 13.5 |
| 10 | 1.8232 | 8 | 75.3 | 1 | 27.7 | 7 | 42.6 | 1 | 01.1 |
| 11 | 1.7849 | 2 | 68.5 | 11 | 54.7 | 9 | 29.7 | 11 | 22.6 |
| 12 | 1.7464 | 7 | 75.2 | 15 | 35.1 | 8 | 43.2 | 3 | 08.6 |
| 13 | 1.7077 | 6 | 74.9 | 3 | 36.1 | 10 | 44.1 | 15 | 14.5 |
| 14 | 1.6688 | 5 | 74.5 | 14 | 52.1 | 13 | 43.1 | 6 | 18.4 |
| 15 | 1.6297 | 4 | 73.7 | 9 | 57.9 | 11 | 41.2 | 9 | 20.2 |

The percentages of remained energy after four steps of replacement for each Ibis in the two considered scenarios are shown in Figure 10.

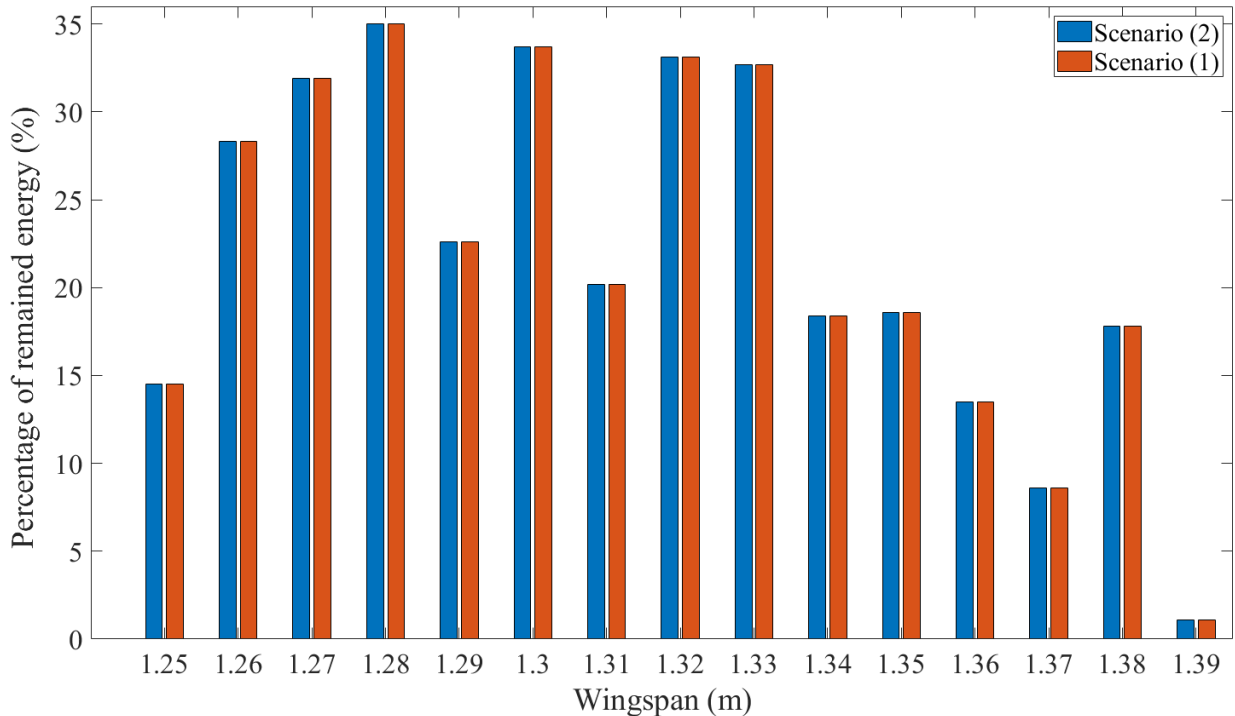

**Figure 10. Percentages of remained energy of Ibises after four steps of replacement for scenarios (1) and (2).**

Figure 10 demonstrates the percentage of the remained energy, which gives a better idea about the level of remained energy for each Ibis based on their initial energy. In Figure 11, the final positions of the Ibises in the flock are shown for two studied scenarios. As can be seen, after four steps of replacement, the two largest birds with wingspans of 1.39 m and 1.37 m have the lowest level of energy. The smallest Ibis with a wingspan of 1.25 m also has consumed more energy than the rest of the Ibises. Moreover, it can be found that small birds have a chance to take the lead position (Ibis with wingspan 1.26 m in scenario 2).

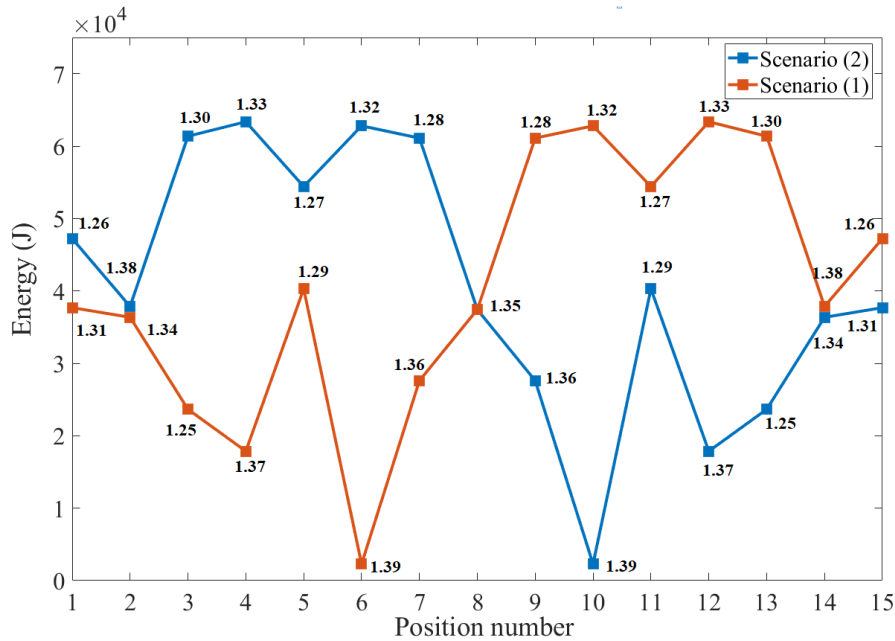

**Figure 11. Final positions of the Ibises in the flock for two studied scenarios.**

## VI.  Conclusion

In this paper, load balancing is conducted on flocking migratory Ibises with different wingspans to increase their flight efficiency. Moreover, it was shown that wingtip spacing plays an important role in the drag of individual birds in the flock. Two different scenarios were considered in this study. In the first scenario, it was assumed that birds in the flock are positioned initially from the smallest to the largest wingspan toward the tail and in the second scenario from the largest to the smallest wingspan. In the repositioning and reformation process, the Ibises which have spent too much energy during the flocking flight were replaced with Ibises that consumed less energy. It was indicated that, after four steps of replacement, Ibises with wingspans of 1.39 m and 1.37 m have the lowest level of remained energy. Furthermore, it was found that small birds have a chance to take the lead position during the flocking flight.

American Institute of Aeronautics and Astronautics